\documentstyle[11pt]{article}         
\begin{document}
\begin{titlepage}
\title{Quantum dynamics with two Planck constants and the
semiclassical limit}
\author{Lajos Di\'osi\thanks{E-mail: diosi@rmki.kfki.hu}\\
KFKI Research Institute for Particle and Nuclear Physics\\
H-1525 Budapest 114, POB 49, Hungary\\\\
{\it bulletin board ref.: quant-ph/9503023}}
\date{March 28, 1995}
\maketitle
\begin{abstract}
The mathematical possibility of coupling two quantum dynamic systems
having two different Planck constants, respectively, is investigated.
It turns out that such canonical dynamics are always irreversible.
Semiclassical dynamics is obtained by letting one of the two Planck
constants go to zero. This semiclassical dynamics will preserve
positivity, as expected, so an improvement of the earlier proposals by
Aleksandrov and by Boucher and Traschen is achieved. Coupling of
quantized matter to gravity is illustrated by a simplistic example.
\end{abstract}
\end{titlepage}

More then 100 years ago, fundamental theory of the physics world was
represented by classical canonical dynamics. Twentieth century brought
curious changes: new quantum dynamics was introduced. The apparatus of
quantum dynamics is rather unusual. Nevertheless, a universal rule of
{\it quantization} has been invented to construct the quantum
counterpart of a given classical canonical dynamics. Classical
canonical coordinates $q_1,q_2,\dots$ and momenta $p_1,p_2,\dots$ must
be considered hermitian operators, and the classical Poisson bracket
\begin{equation}
\{A,B\}_P\equiv \sum_n
\Bigl({\partial A\over\partial q_n}{\partial B\over\partial q_n}
     -{\partial A\over\partial p_n}{\partial B\over\partial p_n}\Bigl)
\label{Pbra}\end{equation}
of dynamic variables $A(q,p),~B(q,p)$ must be replaced by a quantum
bracket:
\begin{equation}
\{A,B\}_Q\equiv -{i\over\hbar}[A,B].
\label{Qbra}\end{equation}
The quantum counterpart of classical Liouville equation of motion
$\dot\rho=\{H,\rho\}_P$ will then be the Schr\"odinger
(or von Neumann) equation
\begin{equation}
\dot\rho=\{H,\rho\}_Q
\label{Sch}\end{equation}
where $H(q,p)$ is the Hamiltonian and $\rho$ is the density operator.

Let us outline the opposite way, i.e., obtaining classical dynamics
from a given quantum one. Basically, one takes the limit
$\hbar\rightarrow0$
and introduces an (overcomplet) basis of normalized states
$\vert q,p\rangle$
for the prospective classical system. These states are wave packets
with center $q,p$. In the limit $\hbar\rightarrow0$, they can and
will be chosen (asymptotically) zero-spread in {\it both} $q$ and $p$.
Taking this basis to work in, all dynamic variables $A(q,p)$ gets
asymptotically diagonal so their diagonal elements will be identified
as the corresponding classical variables:
\begin{equation}
\lim_{\hbar\rightarrow0}\langle q,p\vert A \vert q,p\rangle
=A(q,p).
\label{QACA}\end{equation}
In a similar way the density operator will be assumed (asymptotically)
diagonal in the basis $\vert q,p\rangle$; its diagonal $\rho(q,p)$
corresponds to the phase space distribution of the classical system.
Expanding the quantum commutator (\ref{Qbra}) in the leading order in
$\hbar$ one can prove the asymptotic equation
\begin{equation}
\lim_{\hbar\rightarrow0}\langle q,p\vert\{A,B\}_Q\vert q,p\rangle
=\{A(q,p),B(q,p)\}_P.
\label{QbraPbra}\end{equation}
In this sense can one identify classical counterpart of a given
quantum dynamics by the limit $\hbar\rightarrow0$ of the latter.

Nowadays, quantum dynamics is thought the fundamental one and
classical dynamics is considered a special limit of it. Nevertheless,
there is at least one classical dynamics whose quantization is still
problematic. As a matter of fact, no experimental evidence up to now
has indicated that {\it gravitation} would be a quantum dynamics. And
even theoretical projects of quantizing the classical equations of
gravity have failed to be conclusive enough. So, gravitation could
happen to be classical. Then, the classical dynamics of gravitation
would couple with the quantum dynamics of other fields. This
mathematical problem is not at all trivial. For instance,
Aleksandrov's semiclassical dynamics \cite{Ale} was found \cite{Bou}
to violate trivial conditions of positivity. There is no consistent
theory to couple classical and quantum systems together \cite{And}.

In this Letter, we make an attempt to settle the problem.
Our principal aim is to discuss if classical and quantum canonical
dynamics could be coupled {\it at all\/} in a consistent mathematical
scheme. We would not intend to discuss interpretation of the obtained
results. Yet, we anticipate the basic lesson. It seems unavoidable
that a coupled classical-quantum dynamics be {\it irreversible} which
is in pronounced contrast to the pure classical or pure quantum
canonical dynamics though it is less strange from viewpoint of quantum
measurement theory \cite{Per}.

We start with a system composed of two canonical subsystems and we
assume {\it different} Planck constants for each subsystems so that,
e.g.,
$[q_1,p_1]=i\hbar_1$ and $[q_2,p_2]=i\hbar_2$ with, say,
$\hbar_1>\hbar_2$. In the end, we shall take the limit
$\hbar_1\rightarrow\hbar,~\hbar_2=\rightarrow0$
and in such a way shall we obtain a
(hybrid) {\it semiclassical} dynamics where $(q_1,p_1)$
are quantum and $(q_2,p_2)$ are classical. We prefer this indirect way
in order to utilize a powerful classification of quantum equations of
motion, due to Lindblad \cite{Lin,Gor}. To our knowledge, no
classification is available concerning semiclassical
dynamics directly.

Coming back to the two $\hbar$'s, we emphasize that trading with such
parametric freedom is mathematically trivial and does not lead to any
conflict as long as the states $\rho_n$ of the two systems $(n=1,2)$
evolve independently of each other according to their Schr\"odinger
equations [cf. Eq.~(\ref{Sch})]:
$\dot\rho_n=-(i/\hbar_n)[H_n,\rho_n]$
where $H_n=H_n(q_n,p_n)$ are the corresponding Hamiltonians.

Let us turn to the case of interaction. The total Hamiltonian takes
the form $H=H_1+H_2+H_I$ and the interaction Hamiltonian can in
general be expanded into a series of interacting "currents":
\begin{equation}
H_I(q_1,p_1,q_2,p_2)
=\sum_\alpha J_1^\alpha(q_1,p_1)J_2^\alpha(q_2,p_2),
{}~~~~(J_1^\alpha,J_2^\alpha\ne1).
\label{HI}\end{equation}
In fact, all dynamic variables of the composed system can be
decomposed in a similar form:
$A(q_1,p_1,q_2,p_2)
=\sum_\alpha A_1^\alpha(q_1,p_1)A_2^\alpha(q_2,p_2).$
Hereafter, until the last part of the Letter, I will spare notations
of summation signs and indices so that I write $H_I=J_1J_2$ and
$A=A_1A_2$,  $B=B_1B_2$ e.t.c.\@.

Now, let us find a suitable generalization of the quantum bracket
(\ref{Qbra}) for the case of the composed system with the two Planck
constants. Let us try the ansatz
\begin{equation}
\{A,B\}_Q\equiv -{i\over2\hbar_1}[A_2,B_2]_+ [A_1,B_1]
                  -{i\over2\hbar_2}[A_1,B_1]_+ [A_2,B_2].
\label{genQbra}\end{equation}
Then the Schr\"odinger equation of motion (\ref{Sch}) could be written
in the following form:
\begin{equation}
\dot\rho=-{i\over\hbar_1}[H_1,\rho]-{i\over\hbar_2}[H_2,\rho]
         -{i\over\hbar_{av}}[H_I,\rho]
  +{i\over2}\Delta\hbar^{-1}\left(J_1 \rho J_2 - J_2 \rho J_1 \right)
\label{genSch}\end{equation}
where
$\hbar_{av}=2\hbar_1\hbar_2/(\hbar_1+\hbar_2)$ and
$\Delta\hbar^{-1}=\hbar_2^{-1}-\hbar_1^{-1}$.

The first three terms on the RHS of Eq.~(\ref{genSch}) generate
unitary evolution which is however distorted by the fourth term. It
would not be a problem if it did not violate positivity of $\rho$. But
it does. Seemingly, we should not ask too sharp questions in the
presence of the dynamics $\hbar_1\neq\hbar_2$. We should only inquire
about blurred values of the dynamic variables. Consequently, a certain
smoothening mechanism must be built in {\it by hand}. Let us replace
the interaction Hamiltonian (\ref{HI}) by a noisy one:
\begin{equation}
H_I^{noise}(q_1,p_1,q_2,p_2)=
\Bigl(J_1(q_1,p_1)+\delta J_1(t)\Bigr)
\Bigl(J_2(q_2,p_2)+\delta J_2(t)\Bigr)
\label{HInoise}\end{equation}
where $\delta J_1,\delta J_2$ are {\it classical} noises superposed on
the interacting "currents" $J_1,J_2$. We choose their correlations as
follows:
\begin{equation}
\Bigl\langle\delta J_n(t^\prime)\delta J_n(t)\Bigr\rangle_{noise}
        ={1\over2}h_{av}^2\Delta\hbar^{-1}\lambda_n\delta(t^\prime-t),
        ~~(n=1,2),
\label{corr}\end{equation}
with $\lambda_1\lambda_2=1$ which will be justified later. Assume,
for simplicity's, the two noises $\delta J_1, \delta J_2$ are
independent of each other. The total Hamiltonian becomes noisy:
$H^{noise}=H_1+H_2+H_I^{noise}$.
The blurred dynamics is defined by the Schr\"odinger equation of
motion (\ref{genSch}) {\it averaged} over the noise:
\begin{equation}
\dot\rho=\Bigl\langle \{H^{noise,\rho}\}_Q \Bigr\rangle_{noise}.
\label{Schnoise}\end{equation}
Such noisy Hamiltonians are known to generate typical
double-commutator terms \cite{Gor} so the given $H^{noise}$ yields:
\begin{equation}
\dot\rho=\{H,\rho\}_Q-{1\over4}\Delta\hbar^{-1}
\Bigl(\lambda_2[J_1,[J_1,\rho]~]+\lambda_1[J_2,[J_2,\rho]~]\Bigr).
\label{Schnoise1}\end{equation}
By using the notation $F=\lambda_2^{1/2}J_1-i\lambda_1^{1/2}J_2$
and by substituting the definition (\ref{genQbra}) of the generalized
quantum bracket into the above equation one obtains:
\begin{equation}
\dot\rho=-{i\over\hbar_1}[H_1,\rho]-{i\over\hbar_2}[H_2,\rho]
           -{i\over\hbar_{av}}[H_I,\rho]
        -{1\over4}\Delta\hbar^{-1}
\left(F^\dagger F \rho + \rho F^\dagger F -2F \rho F^\dagger \right),
\label{master}\end{equation}
provided $\lambda_1\lambda_2=1$. This equation belongs to the class of
Lindblad master equations \cite{Lin,Gor} so it can be embedded into an
enlarged unitary dynamics. This mathematical correspondence assures
the consistency of the above master equation though this equation is
{\it not} reversible anymore. Its detailed form is:
\begin{eqnarray}
\dot\rho=-{i\over\hbar_1}[H_1,\rho]-{i\over\hbar_2}[H_2,\rho]
         -{i\over\hbar_{av}}[H_I,\rho]\nonumber\\
         -{1\over2}\Delta\hbar^{-1}
\Bigl(iJ_2 \rho J_1 -iJ_1 \rho J_2+{1\over2}\lambda_2[J_1,[J_1,\rho]~]
        +{1\over2}\lambda_1[J_2,[J_2,\rho]~]\Bigr).
\label{masterdetail}\end{eqnarray}

This master equation will be utilized to construct semiclassical
dynamics. According to what we outlined in the first part, to make the
system $q_2,p_2$ classical we need the limit $\hbar_2\rightarrow0$ and
we must work in the asymptotic basis of zero-spread wave packets
$\vert q_2,p_2\rangle$. Also, we assume the density operator $\rho$
is diagonal in this basis:
\begin{equation}
\rho
=\int \rho(q_2,p_2)\vert q_2,p_2\rangle\langle q_2,p_2\vert dq_2dp_2.
\label{rhoq2p2}\end{equation}
Since all dynamic variables are (asymptotically) diagonal in the same
basis $\vert q_2,p_2\rangle$ one expects Eq.~(\ref{masterdetail})
to preserve the diagonal form (\ref{rhoq2p2}) of $\rho$.

Before taking the limit $\hbar_2\rightarrow0$ we re-scale the
$\lambda-$coefficients:
$\lambda_2=\lambda\hbar_2,~~\lambda_1=\lambda^{-1}/\hbar_2$
otherwise the corresponding terms would diverge.
Perform now the limit $\hbar_2\rightarrow0$ on the diagonal element
$\langle q_2,p_2\vert\dots\vert q_2,p_2\rangle$ of the quantum master
equation (\ref{masterdetail}) and apply the Eq.~(\ref{QbraPbra}) in
it. The resulting semiclassical master equation reads:
\begin{equation}
\dot\rho
=-{i\over\hbar}[H,\rho]+{1\over2}\{H,\rho\}_P -{1\over2}\{\rho,H\}_P
           -{1\over4}\lambda[J_1,[J_1,\rho]]
           +{1\over4}\lambda^{-1}\{J_2,\{J_2,\rho\}_P\}_P
\label{scmaster}\end{equation}
where, in the total Hamiltonian $H(q_1,p_1,q_2,p_2)$, the variables
$q_1,p_1$ are operators while the variables $q_2,p_2$ are numbers.
Obviously, $J_1$ is operator and $J_2$ is number. The object $\rho$
above stands for the diagonal part $\rho(q_2,p_2)$ introduced in
Eq.~(\ref{rhoq2p2}). It is density operator of the quantum subsystem
{\it and} phase space distribution of the classical subsystem. Its
trace over the quantum subsystem's states yields the phase space
distribution $\rho_2(q_2,p_2)$ of the classical subsystem while its
integral over $q_2,p_2$ yields the reduced density operator $\rho_1$
of the quantum subsystem. One expects that the consistency of the
quantum dynamics $\hbar_1\neq\hbar_2$, assured by its Lindblad
structure, survives in the semiclassical limit. Hence we claim that
Eq.~(\ref{scmaster}) is a consistent one and it preserves the
positivity of $\rho$ which was not the case in the earlier proposals
\cite{Ale,Bou,And} to couple classical and quantum canonical dynamics
together.

Let us remember the original form (\ref{HI}) of the interaction
Hamiltonian and restore the hidden indices and signs of summations in
Eq.~(\ref{scmaster}):
\begin{eqnarray}
\dot\rho
=-{i\over\hbar}[H,\rho]+{1\over2}\{H,\rho\}_P-{1\over2}\{\rho,H\}_P
           \nonumber\\
-{1\over4}\sum_{\alpha,\beta}
       \lambda_{\alpha\beta}[J_1^\alpha,[J_1^\beta,\rho]]
+{1\over4}\sum_{\alpha,\beta}
       \lambda^{-1}_{\alpha\beta}\{J_2^\alpha,\{J_2^\beta,\rho\}_P\}_P
\label{scmasterindices}\end{eqnarray}
where $\lambda$ is positive  matrix controlling the statistics of
noises needed for consistency. Note that this semiclassical master
equation might have been derived directly by blurring the Aleksandrov
equation \cite{Ale}:
\begin{equation}
\dot\rho=\Bigl\langle -{i\over\hbar}[H^{noise},\rho]
        +{1\over2}\{H^{noise},\rho\}_P-{1\over2}\{\rho,H^{noise}\}_P
                                                \Bigr\rangle_{noise}
\label{Aleks}\end{equation}
with the noisy Hamiltonian (\ref{HInoise}) of correlations:
\begin{equation}
\Bigl\langle
\delta J_1^\alpha(t^\prime)\delta J_1^\beta(t)\Bigr\rangle_{noise}
        ={1\over2}\lambda_{\alpha\beta}^{-1}\delta(t^\prime-t),~~~~
\Bigl\langle
\delta J_2^\alpha(t^\prime)\delta J_2^\beta(t)\Bigr\rangle_{noise}
        ={\hbar^2\over2}\lambda_{\alpha\beta}\delta(t^\prime-t).
\label{Alekscorr}\end{equation}

Finally, let us see an illustrative example. One might consider simple
models possessing exact solutions (like, e.g., harmonic coupling
between quantum and classical oscillators). Instead of doing so, we
recall the original issue as to couple quantized matter with classical
gravitation. So we apply the semiclassical master equation
(\ref{scmasterindices}) to the interaction of quantized
nonrelativistic matter with weak classical gravitational field. The
quantum subsystem's Hamiltonian is $H_m(q,p)$ with conjugate
variables $q_n,p_n$, $n=1,2,\dots$\@. Let us introduce the Newtonian
potential
$\phi\equiv{1\over2}c^2(g_{00}-1)$
where $g_{00}$ is the relevant component of the metric tensor and $c$
is the velocity of light. We consider the {\it field} $\phi(r)$
canonical coordinates of the gravitational dynamics and we denote its
canonical conjugate momenta by the field $\pi(r)$. The total
Hamiltonian of the interacting system takes the following form:
\begin{equation}
H(q,p,\phi,\pi)=H_m(q,p)
+{1\over8\pi G}
\int_r\left({1\over c^2}\pi^2+\vert\nabla\phi\vert^2\right)
+\int_r f(r)\phi(r)
\label{Hgrav}\end{equation}
where $G$ is the Newton constant and $f(r)$ stands for the mass
distribution {\it operator} of the quantized matter \cite{foot}. By
comparing the third (interaction) term on the RHS of Eq.~(\ref{Hgrav})
to the RHS of Eq.~(\ref{HI}) one identifies the quantized "current"
$J_1^\alpha$ by $f(r)$ and the classical "current" $J_2^\alpha$ by
$\phi(r)$ while, obviously, summations over label $\alpha$ will be
replaced by integrations over the spatial coordinate $r$.

Let us substitute the Hamiltonian (\ref{Hgrav}) into the semiclassical
master equation (\ref{scmasterindices}):
\begin{eqnarray}
\dot\rho=-{i\over\hbar}[H_m,\rho]
                    -{i\over\hbar}\int_r \phi(r)[f(r),\rho]
\nonumber\\
-{1\over4\pi G}\int_r\Bigl(
        {1\over c^2}\pi(r){\delta\rho\over\delta\phi(r)}
        +\Delta\phi(r){\delta\rho\over\delta\pi(r)}\Bigr)
+{1\over2}\int_r \Bigl[f(r),{\delta\rho\over\delta\pi(r)}\Bigr]_+
\nonumber\\
-{1\over4}\int_r\int_{r^\prime}\lambda(r,r^\prime)
        [f(r),[f(r^\prime),\rho]~]
+{1\over4}\int_r\int_{r^\prime}\lambda^{-1}(r,r^\prime)
        {\delta^2\rho\over\delta\pi(r)\delta\pi(r^\prime)}.
\label{gravmast}\end{eqnarray}
Remember that, in fact, $\rho$ stands for $\rho(\phi,\pi)$ which is a
hybrid of density operator for matter and of phase space distribution
for gravity. For instance, its functional integral yields the reduced
density operator $\rho_m$ of the quantized matter:
$\rho_m=\int\int\rho(\phi,\pi){\cal D}\phi{\cal D}\pi$.
We concentrate on the reduced dynamics of $\rho_m$. In addition to
assuming weak gravity, Newtonian approximation will be considered so
that we neglect the term with factor $1/c^2$ on the RHS of
Eq.~(\ref{gravmast}) and suppose that matter's quantum state $\rho_m$
determines Newton potential $\phi$ via the following ansatz:
\begin{equation}
\int \rho(\phi,\pi){\cal D}\pi=
\prod_r\delta\Bigl(
        \phi(r)+\int_{r^\prime} {G/2\over\vert r-r^\prime\vert}
                       [f_+(r^\prime)+f_-(r^\prime)]\Bigr)\rho_m.
\label{rhophi}\end{equation}
The subscripts $+$ and $-$ indicate that the operator $f$ is to
multiply $\rho_m$ from the left or from the right, respectively.
[In such a way hermiticity of $\rho$ is retained by
Eq.~(\ref{rhophi}).] Let us integrate both sides of
Eq.~(\ref{gravmast}) over the fields $\phi,\pi$ while we substitute
the ansatz (\ref{rhophi}) into it. We obtain the following result:
\begin{equation}
\dot\rho_m=-{i\over\hbar}[H_m+H_g,\rho_m]
-{1\over4}\int_r\int_{r^\prime}\lambda(r,r^\prime)
        [f(r),[f(r^\prime),\rho_m]
\label{gravmast1}\end{equation}
where $H_g$ is the well-known Newtonian potential energy:
\begin{equation}
H_g=-{G\over2}\int_r\int_{r^\prime}
{f(r)f(r^\prime)\over\vert r-r^\prime\vert}.
\label{Newton}\end{equation}

One determines the correlation function $\lambda$ by intuitive
considerations. Invoke the interpretation (\ref{Alekscorr}). It
follows that $\lambda$ is related to the hypothetical fluctuations of
$f$ and $\phi$:
\begin{eqnarray}
\Bigl\langle\delta f(r^\prime,t^\prime)\delta f(r,t)
                                                \Bigr\rangle_{noise}
={1\over2}\lambda^{-1}(r^\prime,r)\delta(t^\prime-t),\nonumber\\
\Bigl\langle\delta\phi(r^\prime,t^\prime)\delta\phi(r,t)
                                                \Bigr\rangle_{noise}
        ={\hbar^2\over2}\lambda(r^\prime,r)\delta(t^\prime-t).
\label{Alekscorrgrav}\end{eqnarray}
In Newtonian approximation, the expectation values of the Newton
potential $\phi$ and the mass distribution $f$ are related by the
Poisson equation
$\Delta \langle\phi(r)\rangle=4\pi G\langle f(r) \rangle$ as can be
seen easily from Eq.~(\ref{rhophi}). {\it If\/} the fluctuating
"currents", too, satisfied the Poisson equation then
Eqs.~(\ref{Alekscorrgrav}) would lead to the constraint
$\Delta\Delta^\prime\lambda(r,r^\prime)
=(4\pi G)^2\lambda^{-1}(r,r^\prime)$.
The unique translation invariant correlation satisfying this
constraint is
$\lambda(r,r^\prime)=({G/\hbar})\vert r-r^\prime \vert^{-1}$
\cite{DioLuk}. So the reduced master equation (\ref{gravmast}) of the
quantized matter takes the form:
\begin{equation}
\dot\rho_m=-{i\over\hbar}[H_m+H_g,\rho_m]
-{1\over4}\int_r\int_{r^\prime}{G/\hbar\over\vert r-r^\prime \vert}
        [f(r),[f(r^\prime),\rho_m].
\label{Diomast}\end{equation}

This equation represents a simplistic model of semiclassical gravity.
The equation itself was first obtained as a result of heuristic
efforts to originate macroscopic decoherence from gravitational
fluctuations \cite{Dio}. (Its measurement theoretical aspects are
discussed, e.g., in Refs.~\cite{GhiBalPea}.)

To conclude our Letter we mention less fundamental applications of the
proposed semiclassical equation (\ref{scmasterindices}),
namely for interacting quantum
systems whose particular subsystems behave classically at certain
special conditions. In each case, the choice of noises is not unique.
The issue can be fixed by invoking dimensional, symmetry, or other
intuitive considerations (like in the example above). On the other
hand, we guess that even a larger class of (non-white) noises should
be taken into consideration for instance in the relativistic regime
which is completly beyond the scope of our Letter.
\bigskip

This work was supported by the grants OTKA No. 1822/91 and T016047.


\begin{thebibliography}{99}
\bibitem{foot} Mass distribution operator would take the form
        $f(r)=\sum_n M_n\delta(r-q_n)$
in case of pointlike consituents of masses $M_n$. In fact, the
delta-functions are replaced by functions of finite widths according
to the particle's extensions.
\end{thebibliography}
\end{document}